%% file: main.tex
\documentclass[12pt]{article}

\usepackage{graphicx}
\usepackage{xcolor}
\usepackage{comment}
\usepackage{scicite}

\usepackage{times}

\newenvironment{sciabstract}{%
\begin{quote} \bf}
{\end{quote}}

\newcounter{lastnote}

\title{Dynamics of Algorithmic Content Amplification on TikTok}

\author{Fabian Baumann$^{1, 2, \ast}$, Nipun Arora$^{1}$, Iyad Rahwan$^{1,\ast}$, Agnieszka Czaplicka$^{1,3,\ast}$\\\\
\normalsize{$^{1}$Center for Humans and Machines, Max Planck Institute for Human Development}\\
\normalsize{$^{2}$Department of Biology, University of Pennsylvania}\\
\normalsize{$^{3}$Faculty of Physics, Warsaw University of Technology}\\
\normalsize{$^\ast$ Correspondence: fabian.olit$@$gmail.com, rahwan$@$mpib-berlin.mpg.de, agaczapl$@$gmail.com}}

\date{}
\begin{document} 


\baselineskip24pt

\maketitle 
\begin{sciabstract}

Intelligent algorithms increasingly shape the content we encounter and engage with online. TikTok's "For You" feed exemplifies extreme algorithm-driven curation, tailoring the stream of video content almost exclusively based on users' explicit and implicit interactions with the platform. 
Despite growing attention, the dynamics of content amplification on TikTok remain largely unquantified.  
How quickly, and to what extent, does TikTok’s algorithm amplify content aligned with users' interests?  
To address these questions, we conduct a sock-puppet audit, deploying bots with different interests to engage with TikTok’s "For You" feed. 
Our findings reveal that content aligned with the bots' interests undergoes strong amplification, with rapid reinforcement typically occurring within the first 200 videos watched.  
While amplification is consistently observed across all interests, its intensity varies by interest, indicating the emergence of topic-specific biases.
Time series analyses and Markov models uncover distinct phases of recommendation dynamics, including persistent content reinforcement and a gradual decline in content diversity over time.  
Although TikTok’s algorithm preserves some content diversity, we find a strong negative correlation between amplification and exploration: as the amplification of interest-aligned content increases, engagement with unseen hashtags declines. 
These findings contribute to discussions on socio-algorithmic feedback loops in the digital age and the trade-offs between personalization and content diversity.    
\end{sciabstract}

%todo todo todo todo todo todo todo todo todo todo todo todo todo todo todo todo todo todo todo todo todo todo todo todo

% - anonymizing the data, especially with respect to the usernames

\section{Introduction}  

Artificial Intelligence (AI) is playing an increasingly significant role in mediating the information we engage with \cite{wagner_measuring_2021}. From chatbots powered by large language models to content feeds on platforms like Facebook, X, or TikTok, AI shapes the visibility \cite{conti_revealing_2024}, dissemination \cite{bartley_auditing_2021}, and consumption of online content \cite{milli_engagement_2024}. This growing influence of algorithms on the distribution of information can profoundly affect cognitive processes at the individual level, including emotional states and decision-making \cite{cui_effects_nodate,guess_how_2023}. Beyond individual experiences, AI-mediated learning may also transform the emergence and evolution of cultural artifacts  \cite{brady2023algorithm,czaplicka_mutual_2024}. From art \cite{epstein_art_2023} to scientific innovation \cite{sourati_accelerating_2023}, AI not only shapes the spread of existing cultural artifacts \cite{farrell_large_2025} but also shapes the creation of new ones, making the study of AI-driven information mediation a crucial frontier in understanding society’s transition into the digital age \cite{brinkmann_machine_2023,acerbi2019cultural}.

\begin{figure}[!h]
    \centering
    \includegraphics[width=0.99\linewidth]{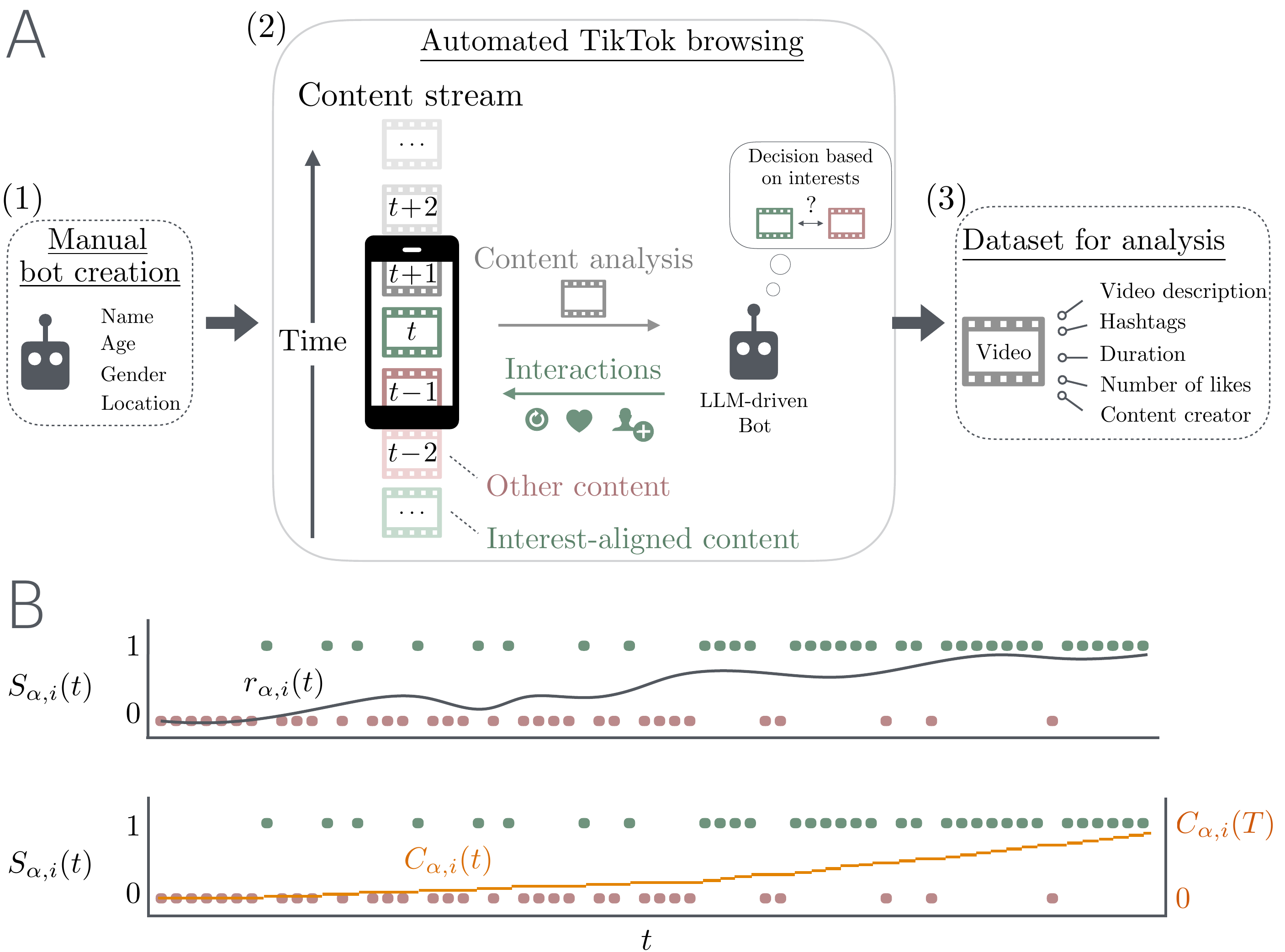}
    \caption{\textit{Sock-puppet audit and content stream analysis.}  
\textbf{Panel A} provides a schematic overview of the sock-puppet audit methodology. The left side illustrates the stream of video content that bots scroll through, where green and red represent interest-aligned and non-aligned videos, respectively. Time is measured as the discrete number of videos, $t \in \{1, \dots, N\}$, with $t=1$ corresponding to the first video a bot encounters at the start of an experimental run. Bots analyze each video by extracting its meta-information (description and hashtags) and processing it through a large language model, which determines whether the content aligns with the bot's predefined interests. If a video matches the bot’s interests, the bot interacts with the platform in three ways: (i) rewatching the video, (ii) liking the video, and (iii) following the video's creator.  
\textbf{Panel B} illustrates how {key quantities for our analysis} are derived from the binary content stream, which encodes whether a video at time $t$ matches the bot’s interest ($S_{\alpha, i}(t) = 1$) or not ($S_{\alpha, i}(t) = 0$). The upper plot schematically depicts the rate of interest-aligned videos over time, $r_{\alpha, i}(t)$, obtained by convolving $S_{\alpha, i}(t)$ with a uniform kernel. 
The lower plot illustrates the cumulative count of interest-aligned videos, $C_{\alpha, i}(t)$, highlighting its monotonically increasing nature and its relationship to $S_{\alpha, i}(t)$. }
\label{fig:overview}
\end{figure}

TikTok represents an extreme case of an algorithm-driven ecosystem \cite{guinaudeau_fifteen_2022,Narayanan2022}. As a short-form video-sharing platform, TikTok allows users to create, share, and engage with highly viral, user-generated content across a wide range of topics, including entertainment, lifestyle, education, and politics. 
Unlike other digital platforms, where content search and discovery is often guided by user queries or influenced by social connections among users, TikTok’s content exposure is almost entirely dictated by its algorithm.
Its "For You" feed curates each user's content diet primarily based on their explicit and implicit interactions with previously encountered videos, such as rewatching, liking, or following content creators. 
This direct feedback loop enables the algorithm to rapidly and continuously adapt to user preferences and emerging trends.

TikTok’s rapid global expansion, combined with its highly personalized user experience, raises critical questions about the societal impact of its algorithm.
Previous studies have examined various aspects of TikTok’s algorithmic feed, including its potentially addictive nature \cite{qin_addiction_2022,yang_studying_2025}, its influence on social connectedness \cite{taylor_lonely_2024}, and the role of user-algorithm feedback loops in shaping content exposure \cite{boeker_empirical_2022,vombatkere_tiktok_2024}.  
Despite these insights, it remains largely unclear how---and to what extent---TikTok’s recommendation algorithm shapes the content distribution individuals encounter by amplifying content aligned with their interests. 

{To address this gap, we conduct a sock-puppet audit by deploying automated bots that interact in real-time with TikTok's "For You" feed, each programmed with distinct content preferences. Extending beyond previous sock-puppet methodologies \cite{habib_youtube_2025,bandy_problematic_2021,mousavi_auditing_2024,bouchaud_auditing_2024}, our approach leverages a large language model (GPT-3.5 Turbo) to dynamically assess video relevance based on descriptions and hashtags. Specifically, bots evaluated each video encountered to determine whether it matched their assigned interests. Upon detecting alignment, the bots performed predefined user-like interactions—such as rewatching, liking, and following the video's creator—to explicitly signal their preferences to the platform's recommendation algorithm.}

Investigating content amplification on TikTok requires selecting topic categories that are sufficiently prevalent to ensure that the bots can effectively signal their interests within a reasonable timeframe.  
To this end, we analyze three experimental conditions: (i) bots with a single interest in \textsc{Gaming} content (condition $G$), (ii) bots with a single interest in \textsc{Food} content ($F$), {and (iii) dual-interest bots engaging with both \textsc{Gaming} and \textsc{Food} content ($G+F$). The dual-interest bots interact with content classified as either \textsc{Gaming}, \textsc{Food}, or both.} See Fig.~\ref{fig:overview} for a schematic overview of our methodology and the Methods section for details on the implementation of the sock puppet audit.

Adopting this user-centric approach enables us to systematically examine how TikTok’s {feed} algorithm shapes content distribution based on user interests.  
{Specifically, we investigate how quickly the recommendation algorithm detects a bot's interests and how it adapts over time.  
Additionally, we analyze how the dynamics of content amplification vary across different bot interests and how amplification correlates with other content characteristics, such as popularity and duration.  
Finally, we explore the relationship between content amplification and content diversity, assessing whether increased reinforcement of interest-aligned content narrows exposure to new topics and perspectives.}  

The paper is organized as follows. In the next section, we present our results, detailing the dynamics of interest-aligned content amplification on TikTok's "For You" feed. We analyze these dynamics using time series analysis and (hidden) Markov models to uncover distinct recommendation patterns. Following this, we discuss our findings in the context of existing research on algorithmic feeds and outline the limitations of our study. {A detailed description of our sock-puppet audit and analysis can be found in the Methods section and the Supplementary Material (SM).}

% interesting references
%\item The Dynamics of Exploration on Spotify \cite{mok_dynamics_nodate}

\section{Results}
\begin{figure}[!h]
    \centering
    \includegraphics[width=0.99\linewidth]{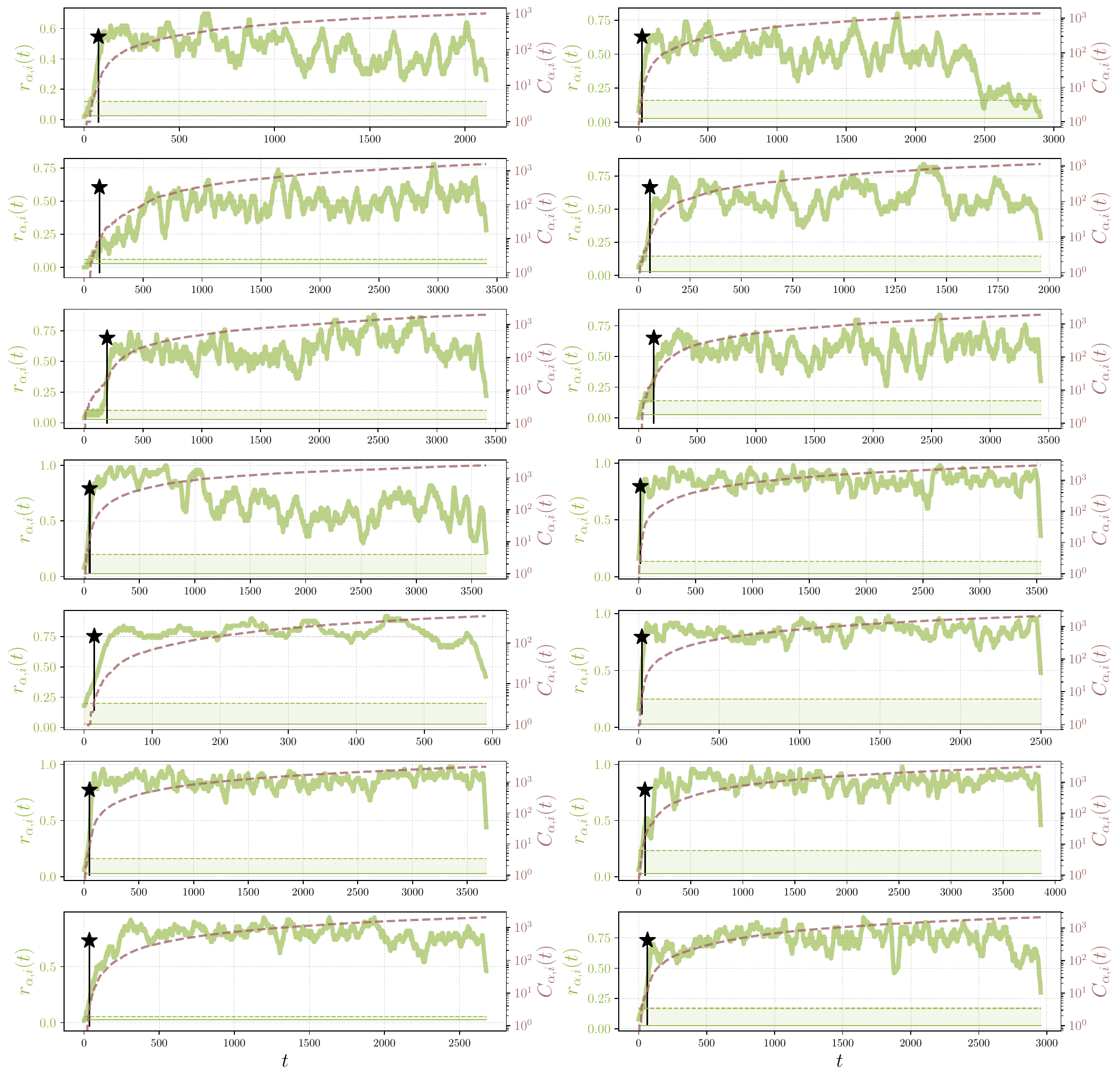}
    \caption{\textit{Content streams for bots interested in \textsc{Gaming} content.}
    The solid green line represents the rate of interest aligned content $r_{G, i}(t)$ and the dashed red line corresponds to the cumulative count of interest-aligned videos $C_{G, i}(t)$, both derived from $S_{G, i}(t)$. The green band spans between the two baselines $b_1$ (dashed line) and $b_2$ (dotted line).
    }
\label{fig:dynamics_gaming}
\end{figure}

\begin{figure}[!h]
    \centering
    \includegraphics[width=0.99\linewidth]{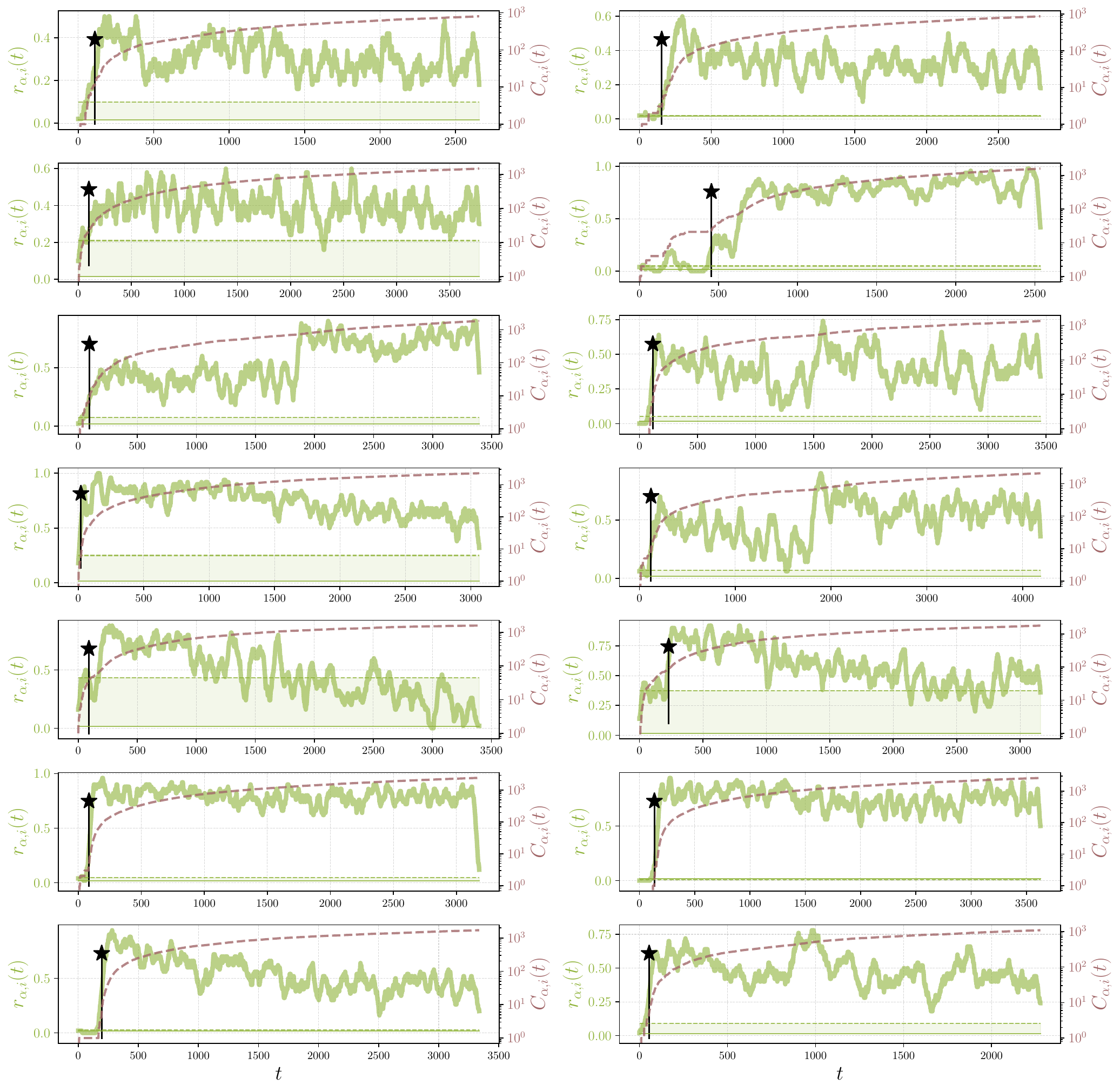}
    \caption{\textit{Content streams for bots interested in \textsc{Food} content.}
    The solid green line represents the rate of interest aligned content $r_{F, i}(t)$ and the dashed red line corresponds to the cumulative count of interest-aligned videos $C_{F, i}(t)$, both derived from $S_{F, i}(t)$. The green band spans between the two baselines $b_1$ (dashed line) and $b_2$ (dotted line). 
    }
\label{fig:dynamics_food}
\end{figure}

\begin{figure}[!h]
    \centering
    \includegraphics[width=0.99\linewidth]{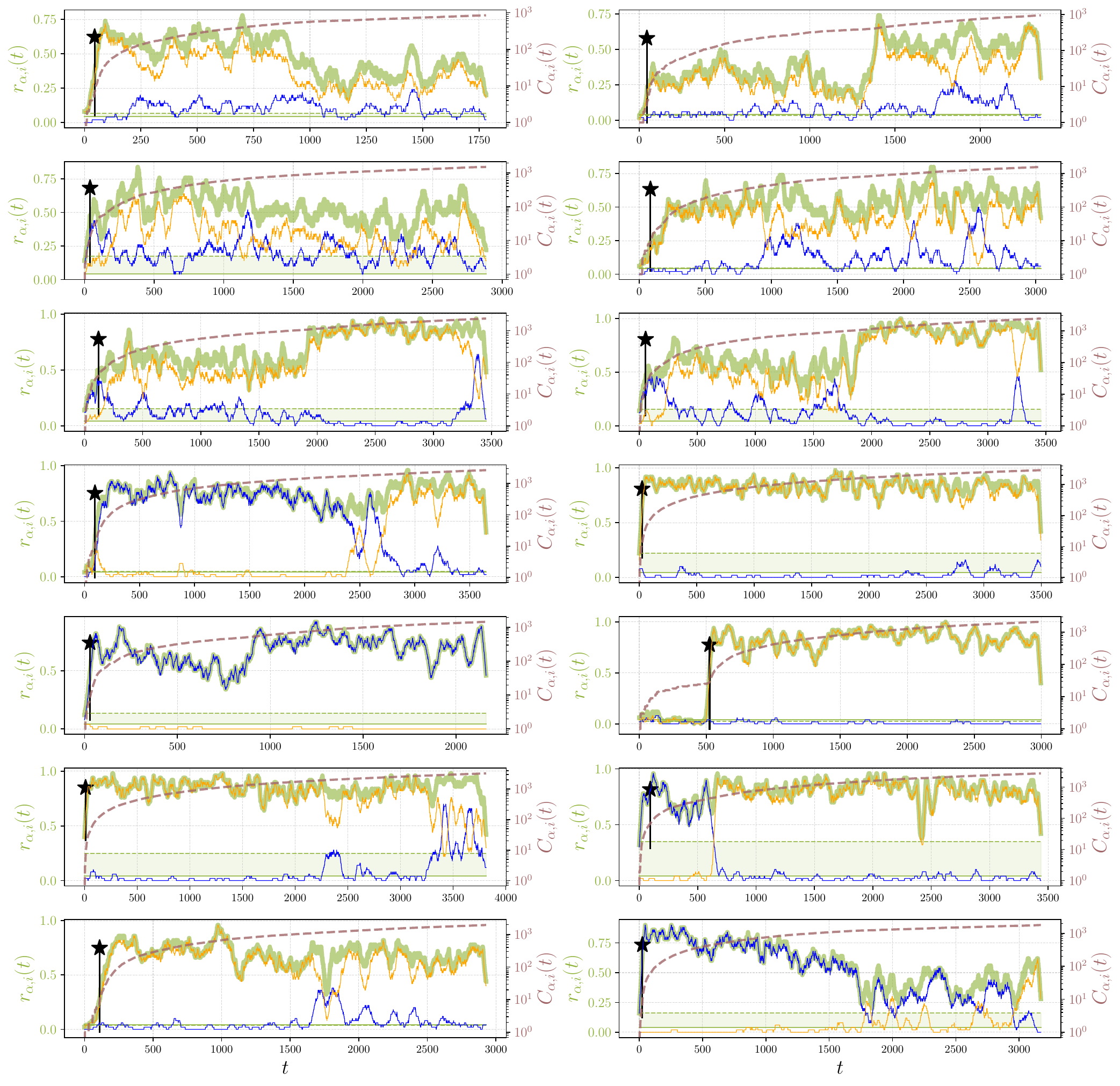}
\caption{{\textit{Content streams for bots interested in \textsc{Gaming} and \textsc{Food} content.}  
The solid green line represents the rate of interest-aligned content, $r_{GF, i}(t)$, while the dashed red line denotes the cumulative count of interest-aligned videos, $C_{GF, i}(t)$, both derived from $S_{GF, i}(t)$.
For dual-interest bots, $S_{GF, i}(t)$ is computed using a logical OR operation, meaning $S_{GF, i}(t) = 1$ if the video at time $t$ belongs to either \textsc{Gaming} or \textsc{Food}, or both categories; otherwise $S_{GF, i}(t) = 0$. The orange (blue) line depicts the rate of \textsc{Gaming} (\textsc{Food}) content.
The green band spans the range between the two baseline frequencies: $b_1$ (dashed line) and $b_2$ (dotted line), where the baseline $b_2$ is computed as the average of the corresponding baselines of the single interest bots.}} 

\label{fig:dynamics_gaming_food}
\end{figure}

\paragraph{{Content Amplification.}}
The dynamics of {video} recommendations for single-interest bots are depicted in Fig.~\ref{fig:dynamics_gaming} (\textsc{Gaming}, G) and Fig.~\ref{fig:dynamics_food} (\textsc{Food}, F), and for dual-interest bots in Fig.~\ref{fig:dynamics_gaming_food} (\textsc{Gaming}+\textsc{Food}, GF). 
{Figures~\ref{fig:dynamics_gaming}--\ref{fig:dynamics_gaming_food} each correspond to one of the three experimental conditions, with each panel depicting the dynamics of one of the twelve (12) experimental realizations (i.e., bots) within that condition.}

{The data from each experiment is represented as a discrete-time series, $S_{\alpha, i}(t)$, where $t \in \{0,1,\dots,T\}$ denotes the index of videos watched by a bot. Thus, the total number of videos in an experiment is given by $T$. The letter $\alpha$ specifies the experimental condition ($\alpha \in \{G, F, G\!+\!F\}$), and $i$ identifies the experimental realization ($i \in \{1,\dots,12\}$).  
For example, $S_{G,3}(t)$ represents the time series of the third experimental realization in the \textsc{Gaming} condition. 
If the 5$^{\mathrm{th}}$ video in the experiment matches the bot's interest (i.e., \textsc{Gaming}), then $S_{G,3}(5) = 1$; otherwise, $S_{G,3}(5) = 0$.}
The frequency of interest-matching videos over time $r_{\alpha, i}(t)$ is approximated as the time-moving average of $S_{\alpha, i}(t)$, i.e. the discrete convolution \( r_{\alpha,i}(t) = \sum_{k=-\lfloor w/2 \rfloor}^{\lfloor w/2 \rfloor} \frac{1}{w} S_{\alpha,i}(t - k) \), where \( w \) is the window size of the averaging kernel. In Figs.~\ref{fig:dynamics_gaming}--\ref{fig:dynamics_gaming_food}, the rate $r_{\alpha, i}(t)$ is depicted as solid green line.

The dashed red lines represent the cumulative count of interest-matching videos encountered by the bot up to time $t$, {given by the cumulative sum  $C_{\alpha, i}(t) = \sum_{t=0}^{t} S_{\alpha, i}(t)$.} {Thus, the total number of interest-aligned videos in an experiment is given by $C_{\alpha, i}(T)$.}
This quantity is displayed on a logarithmic scale on the left y-axis of each plot.

Finally, the {green} band highlights the range between two distinct baseline rates, serving as a reference for comparing the actual rate of interest-matching content, $r_{\alpha, i}(t)$. 
First, the dashed green line represents the baseline $b^{\alpha, i}_1 = C_{\alpha, i}(t_{o}) / t_{o}$, which corresponds to the average rate of interest-aligned content during the \textit{initial phase} of the experiment, up to the onset point $t_{o}$ (star symbols), before any significant amplification of interest-aligned content occurs (see below for details). 
This baseline value varies across experimental realizations $i$.
Second, the solid green line represents the baseline {$b^\alpha_2 = \langle C_{\gamma,j}(T)\rangle_j$ with $\alpha\neq\gamma$,} which denotes the average rate of interest-matching content (e.g., \textsc{Gaming} content) in conditions where the bot primarily has a different interest (e.g., \textsc{Food} content).
The baseline frequency $b^\alpha_2$ does \textit{not} vary across experimental realizations. 
{The baseline $b^{G+F}_2$ for the \textsc{Gaming+Food} condition is calculated as the mean of the single-interest baselines, $b_2^\alpha$, given by
$b^{G+F}_2 = (b^{G}_2 + b^{F}_2)/2$.}

We observe a pronounced and rapid amplification of content aligned with the bots' interests for both single-interest bots (Figs.~\ref{fig:dynamics_gaming} and \ref{fig:dynamics_food}) and dual-interest bots (Fig.~\ref{fig:dynamics_gaming_food}). This amplification is often reflected in the sudden and steep increase of $r_{\alpha, i}(t)$ shortly after the start of an experiment, followed by its sustained elevation at later times compared to both baselines ($b^{\alpha, i}_1$ and $b^\alpha_2$). 

\begin{figure}[!h]
    \centering
    \includegraphics[width=0.95\linewidth]{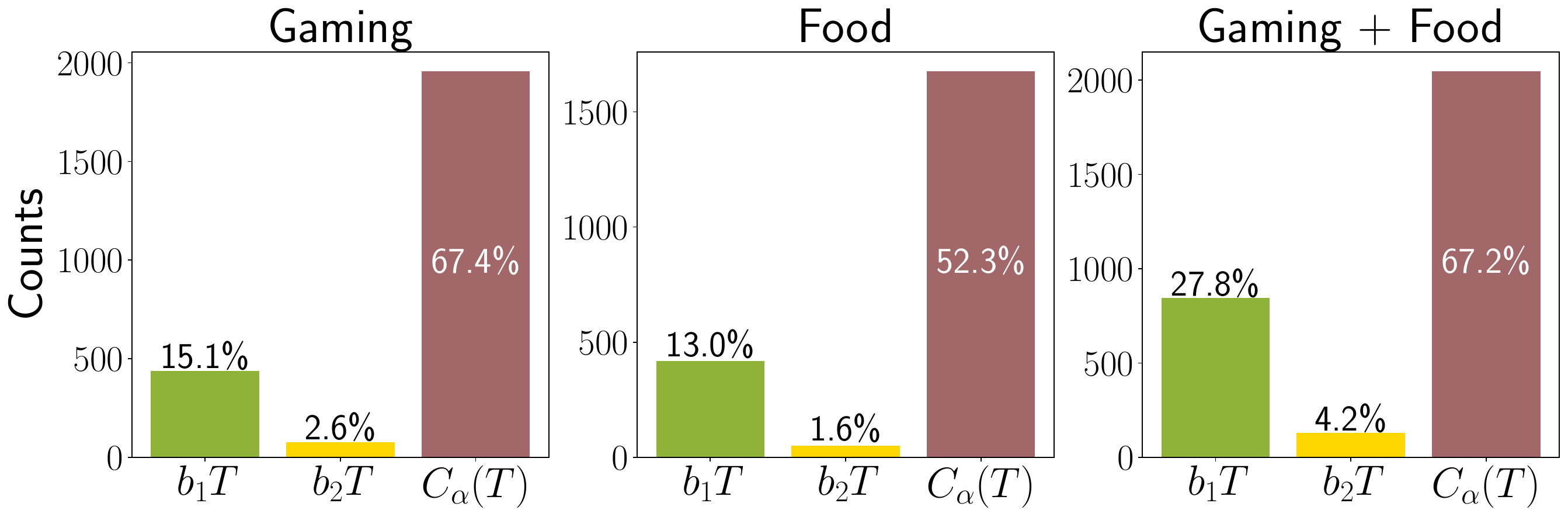}
    \caption{\textit{Cumulative counts of interest-aligned videos.}  
    The dark red bars represent the actual number of interest-aligned videos encountered by the bot, compared to the expected counts derived from the two baseline frequencies, $b_1$ and $b_2$. The percentages indicate the proportion of interest-aligned videos relative to the total number of videos watched.}  
\label{fig:cumulative_counts}
\end{figure}

The strong amplification of interest-aligned content is further evident when comparing the average cumulative count of interest-aligned videos in each condition, {defined as $C_\alpha(T) = \langle C_{\alpha, i} \rangle_i$, to the expected cumulative counts derived from the baselines $b_1$ and $b_2$,} as shown in Fig.~\ref{fig:cumulative_counts}.  
Across all {conditions}, the total number of interest-aligned videos consistently surpasses the baseline expectations $b_1 T$ and $b_2 T$. For single-interest bots, the amplification is most pronounced in the \textsc{Gaming} condition, somewhat lower in the \textsc{Food} condition, and slightly more moderate for dual-interest bots (\textsc{Gaming}+\textsc{Food}).  
Notably, these cumulative counts $C$ correspond to strikingly high ratios of interest-aligned content: $67.4\%$, $52.3\%$, and $67.2\%$ for \textsc{Gaming}, \textsc{Food}, and \textsc{Gaming}+\textsc{Food}, respectively.

\paragraph{Onset of Content Amplification.}
We estimate the time at which TikTok's feed algorithm identifies the bots' interests as the point when it begins significantly increasing recommendations of interest-aligned content. To determine this, we apply an established change point detection method \cite{truong2020selective}.  
In Figs.~\ref{fig:dynamics_gaming}, \ref{fig:dynamics_food}, and \ref{fig:dynamics_gaming_food}, the onset times $t_{o}$ are indicated by vertical black lines and star symbols.

\begin{figure}[!h]
    \centering
    \includegraphics[width=0.95\linewidth]{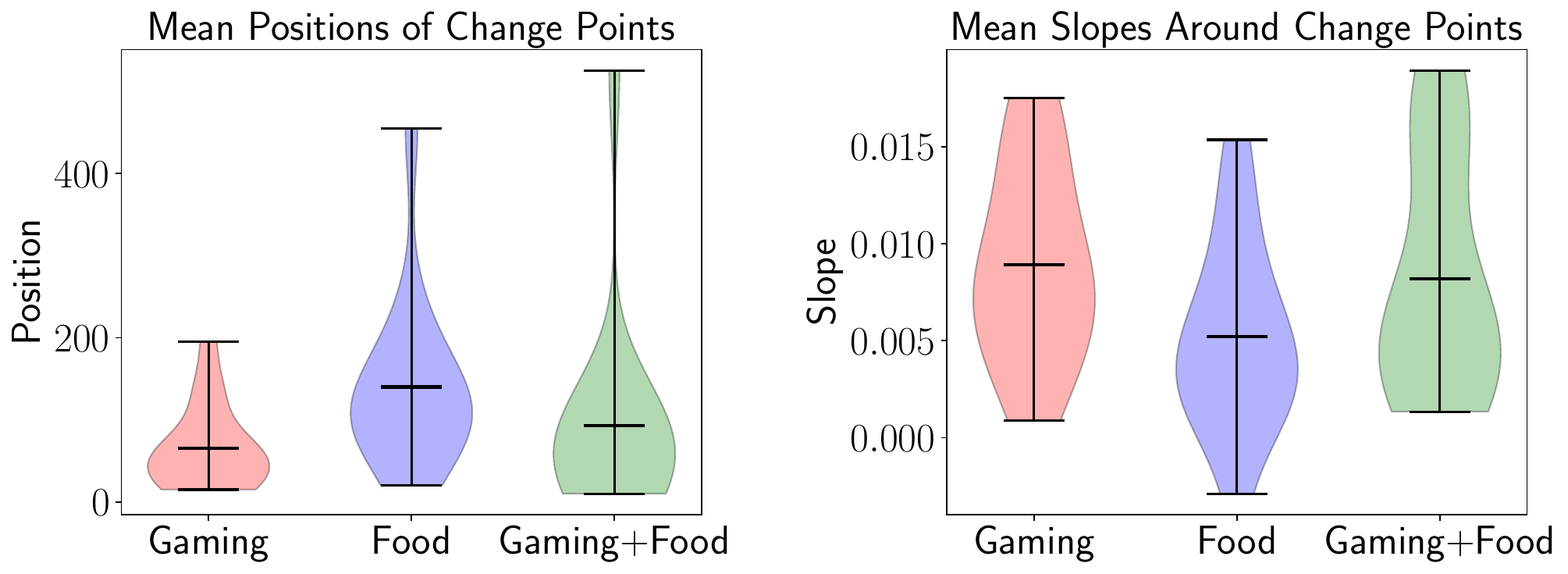}
    \caption{\textit{Onset of content amplification.} The left panel shows the positions of the onset change points, while the right panel depicts the slopes of content amplification for the three conditions (\textsc{Gaming}, \textsc{Food}, \textsc{Gaming}+\textsc{Food}). Each violin plot represents the distribution of data from 14 runs per condition. The horizontal lines inside the violins indicate the mean values, and the extents of the violin outline reflect the data distribution. The upper and lower edges of the vertical lines correspond to the minimum and maximum observed values.}
\label{fig:onset_stats}
\end{figure}

The onset points are determined from the binary time-series {$S_{\alpha, i}(t)$ and are typically located in regions where $r_{\alpha, i}(t)$} exhibits a sharp initial increase.  
Notably, most onset points occur within the first 200 videos viewed by the bots, though variations exist across experimental conditions, as depicted in Fig.~\ref{fig:onset_stats}. On average, the onset occurs after approximately 65.7 videos for single-interest \textsc{Gaming} bots, 140 videos for single-interest \textsc{Food} bots, and 93.2 videos for dual-interest \textsc{Gaming}+\textsc{Food} bots.

After identifying the recommendation onset, we can quantify how rapidly early content amplification unfolds. Specifically, we estimate this {aspect of amplification by calculating the slope $m_{\alpha, i}$ of $r_{\alpha, i}(t_o)$ around the onset points $t_o$, given by $m_{\alpha, i} = [r_{\alpha, i}(t_o + d) - r_{\alpha, i}(t_o - d)] / 2d$, where $d$ represents the window size, i.e., the number of time steps (subsequent videos) used to compute the slope.} In the SM the linear fits are depicted for each experimental condition.

We find that the slopes $m_{\alpha, i}$ vary strongly across conditions $\alpha \in \{G, F, GF\}$, as shown in the right panel of Fig.~\ref{fig:onset_stats}. The onset dynamics for \textsc{Gaming} content exhibit the steepest average slope (0.009), while \textsc{Food} content shows the lowest (0.005). For dual-interest bots, the average slope (0.008) falls between those of the two single-interest conditions.

\paragraph{Post-Onset Dynamics.}
To analyze the later-phase behavior of the recommendation dynamics, we first apply a simple Markov model. Under the assumption of stationarity, we derive both the transition probabilities and the stationary probability distributions of the states under the model.

\begin{figure}[!h]
    \centering
    \includegraphics[width=0.99\linewidth]{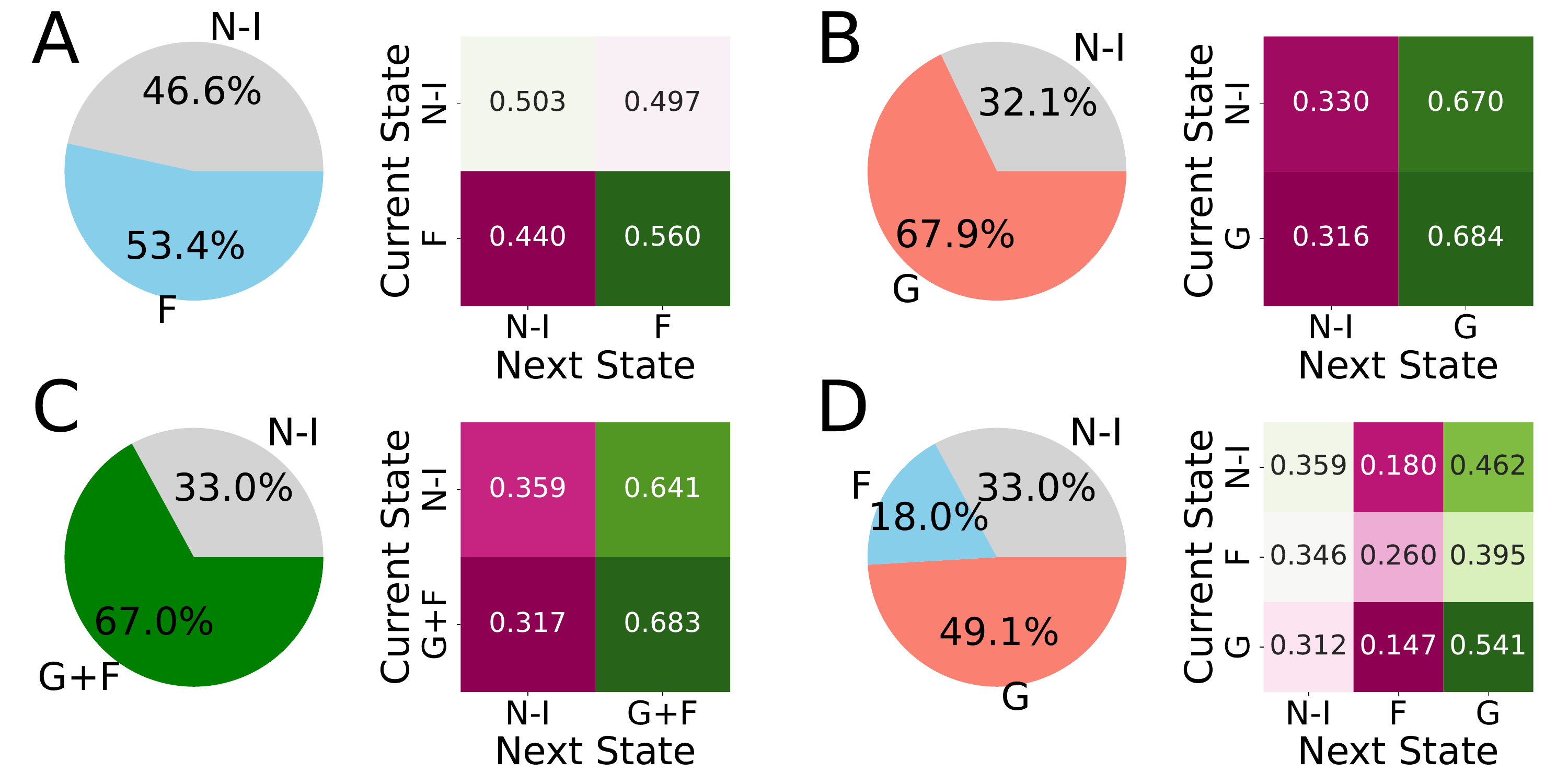}
        \caption{\textit{Stationary Dynamics of Recommendations inferred from Markov model}. Stationary distributions and transition probabilities between recommendation states (non-interest and interest) for single interest bots (\textbf{Panels A, B}) and dual interest bots (\textbf{Panels C, D}).}
\label{fig:markov_model}
\end{figure}

For single-interest bots, the system is modeled using two states: non-interest content (N-I, state 1), and interest-aligned content (\textsc{Gaming}/\textsc{Food}) (state 2). 
For dual-interest bots, we examined two scenarios: (i) a \textit{two-state} Markov model with non-interest content (state 1), and interest-aligned content, which aggregates \textsc{Gaming}- and \textsc{Food} content (state 2), and (ii) a \textit{three-state} Markov model that distinguishes between non-interest content (state 1) and the two interest-aligned content types \textsc{Gaming} (state 2), and \textsc{Food} (state 3), adding complexity but offering a more detailed representation of the content dynamics.

The results of the Markov model are shown in Fig.~\ref{fig:markov_model}. For single-interest bots, the average transition probabilities exhibit a strong dependence on the topic of interest. In the \textsc{Food} condition, following the recommendation of a non-interest video, content aligned with the bots' interest appears  $49.7\%$ of the time. This probability increases to $56\%$ after an interest-aligned recommendation, indicating a higher likelihood of \textsc{Food} content in subsequent recommendations.
Overall, the inferred transition probabilities of the Markov model give rise to an average stationary probability distribution of $46.6\%$ non-interest content (state 1) and $53.4\%$ \textsc{Food} content (state 2), respectively.

For single-interest \textsc{Gaming} bots, the Markov model reveals an even stronger amplification of interest-aligned content. Transitions from both states---non-interest (state 1) and interest (state 2)---exhibit an elevated probability to be followed by interest-aligned content. Specifically, non-interest content transitions to interest-aligned content with a probability of $67\%$, and interest-aligned content is followed by further interest-aligned content with an even higher probability of $68.4\%$. The average stationary distribution in the \textsc{Gaming} condition evaluates to $32.1\%$ non-interest content (state 1) and $67.9\%$ interest-aligned \textsc{Gaming} content (state 2).

The Markov model also reveals a strong amplification of interest-aligned content for the dual-interest bots. When aggregating the two interests (\textsc{Gaming} and \textsc{Food}) into a single state, we observe that interest-aligned content follows non-interest content in $64.1\%$ of cases and follows interest-aligned content in $68.3\%$ of cases, resulting in a stationary probability distribution of $33\%$ for non-interest content and $67\%$ for interest-aligned content.

\begin{figure}[!h]
    \centering
    \includegraphics[width=0.99\linewidth]{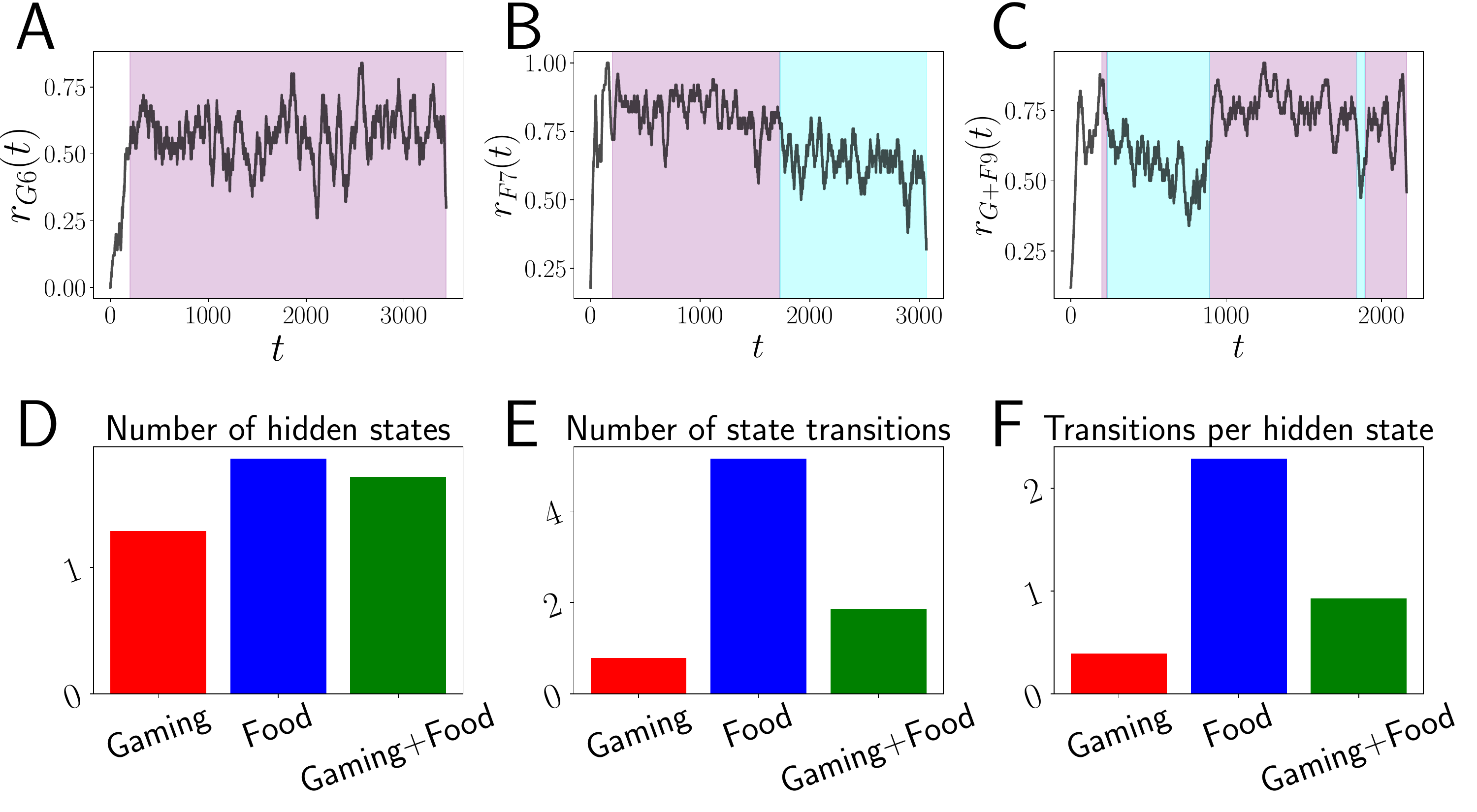}
    \caption{\textit{Results of the Hidden Markov Model.} Examples of the recommendation dynamics with hidden states from different experimental conditions (\textbf{Panels A-C}) and average results of the Hidden Markov Model across conditions (\textbf{Panels D-F}).}
\label{fig:hidden_markov_modeling}
\end{figure}

The three-state Markov model allows for a more detailed analysis, giving insights into which topic might mostly drive the content amplification in the dual-interest condition.
Indeed, the bottom-right panel of Fig.~\ref{fig:markov_model} suggests that \textsc{Gaming} content accounts for most of the amplification of interest-aligned content. 
Moreover, the transition probabilities reveal that transitioning to \textsc{Gaming} content is consistently the highest, regardless of the current state: $46.2\%$ from non-interest content, $39.5\%$ from \textsc{Food} content, and $54.1\%$ from \textsc{Gaming} content. In contrast, \textsc{Food} content follows less frequently, with probabilities of $18\%$ after non-interest content, $26\%$ after \textsc{Food} content, and $14.7\%$ after \textsc{Gaming} content. These transition dynamics result in a stationary distribution of $33\%$ for non-interest content, $18\%$ for \textsc{Food} content, and $49\%$ for \textsc{Gaming} content, which is consistent with the two-state model.

While the simple Markov model provides valuable insights into the system's long-term behavior, it is limited to transitions between observable states, potentially overlooking more complex, hidden dynamics. To address this limitation, we also employ a Hidden Markov Model (HMM), which uncovers latent states and offers further insights into the dynamical regimes driving content amplification following its onset.

Figure~\ref{fig:hidden_markov_modeling} showcases the results of the HMM analysis for a subset of experiments, with one example from each experimental condition. 
In panels A-C, the black solid line represents the rate of interest-aligned content, $r_{\alpha, i}(t)$, while the hidden state dynamics are shown as colored areas, with each color corresponding to a distinct hidden state.
While in the example from the \textsc{gaming} condition, the dynamics are best described by a single hidden state, the examples from the \textsc{food} and \textsc{gaming+food} conditions are best explained by two hidden states. 
Some of the transitions between hidden states are associated with abrupt changes in the rate of interest-aligned content (e.g., panel C), others, instead, reflect smoother shifts of $r_{ij}(t)$ (e.g. panel B).

\begin{figure}
    \centering
    \includegraphics[width=0.99\columnwidth]{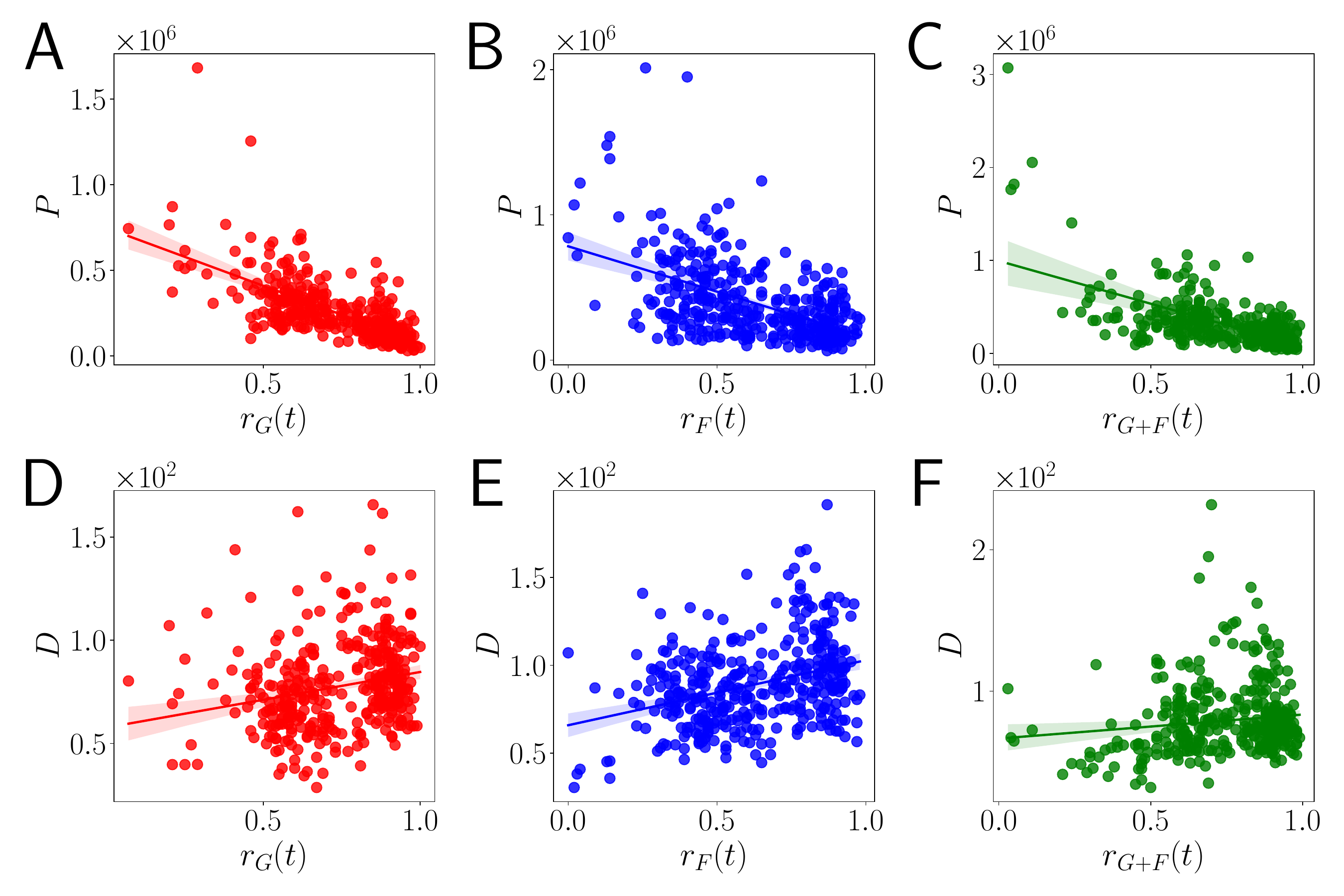}
    \caption{\textit{Video popularity and duration.} \textbf{Panels D-F} show popularity of the videos (measured by the number of 'likes') vs rate of interest-matching content. Points represent results averaged over a moving window of size 100. Panels presents results for bots grouped according to their interests, left panel FOOD bots, center panel GAMING bots and right panel FOOD+GAMING bots. \textbf{Panels D-F} depict the duration of videos (measured in seconds) vs rate of interest-matching content. Points represent results averaged over a moving window of size 100. Panels presents results for bots grouped according to their interests, left panel FOOD bots, center panel GAMING bots and right panel FOOD+GAMING bots.}
    \label{fig:popularity}
\end{figure}

%For dual-interest bots, state transitions can also likely correspond to shifts in the recommender system's focus between the two topics of interest. 
%These transitions can represent either a complete switch in focus (e.g. \fb{X}) or a significant distortion of the topic ratio, where one topic dominates at the expense of the other (e.g., \fb{Y}). 

The lower panels of Fig.~\ref{fig:hidden_markov_modeling} summarize the average results of the inferred hidden states across experimental conditions. 
Panel D shows that the average number of hidden states inferred from {$r_{\alpha, i}(t)$} varies by condition, with the highest value for the \textsc{food} condition and the lowest value for the \textsc{gaming} condition. The dual-interest condition (\textsc{gaming+food}) exhibits an intermediate average number of hidden states.
This variation between conditions becomes even more pronounced when examining the number of hidden state \textit{transitions} (panel E) and the number of transitions normalized by the number of hidden states (panel F). The \textsc{food} condition shows a much higher mean number of state transitions compared to both the \textsc{gaming} and \textsc{gaming+food} conditions, highlighting the increased dynamical complexity in the \textsc{food} condition. 
%\fb{\textbf{[check if both panels E, F are needed. If not consider using sth like entropy to quantify the distribution of the signal across hidden states]}}

\paragraph{Video duration and popularity.}  
{We examined how video duration and popularity vary with the degree to which content aligns with the bots' interests, as quantified by $r_{\alpha, i}(t)$}.

Panels A-C in Fig.~\ref{fig:popularity} present the results for video popularity $P$, which we approximate by the number of likes a video has received at the time it was viewed by the bot. 
Across all experimental conditions, we observe a clear \textit{negative} correlation between popularity and the rate of interest-aligned content (x-axis): as content becomes more aligned with the bot's interests, its popularity decreases.  
Panels D-E summarize the findings for video duration $D$, where we observe the opposite trend. 
Specifically, across all three conditions, there is a strong \textit{positive} correlation between $r_{\alpha, i}(t)$ and D---content that more closely matches the bot's interests tends to be longer {than more general content.}

\paragraph{Exploration of content.}  To quantify how the amplification of interest-aligned videos affects content exploration, we analyzed the composition of hashtags encountered by the bots over time across the different conditions.  

\begin{figure}
    \centering
    \includegraphics[width=\linewidth]{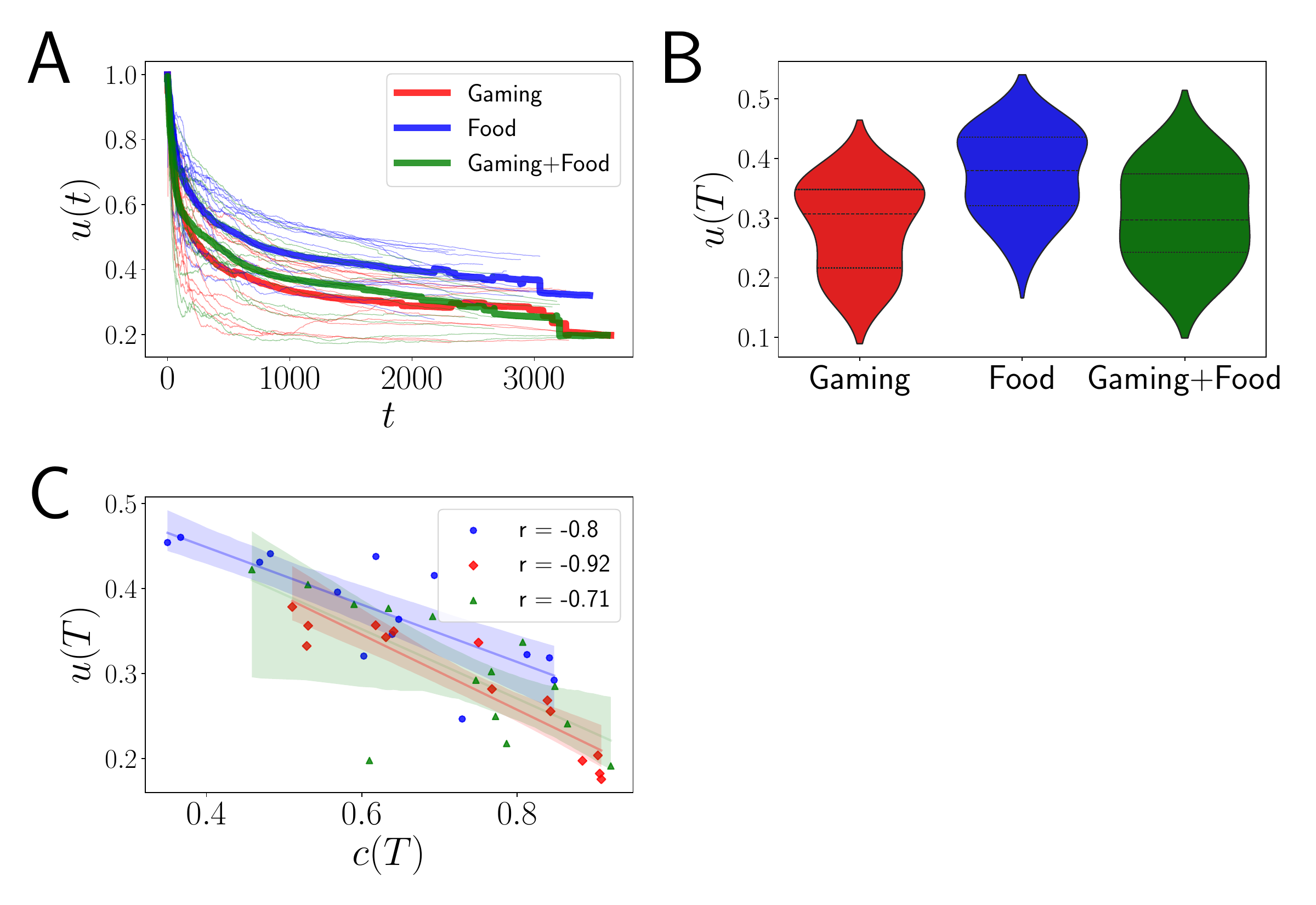}
    \caption{\textit{Exploration of hashtags.}  
    \textbf{Panel A} displays the ratio of unique to total hashtags encountered by each bot across experimental realizations (thin lines) and the corresponding averages for each experimental condition (thick lines).  
    \textbf{Panel B} presents violin plots illustrating the distributions of final unique-to-total hashtag ratios for each condition.  
    \textbf{Panel C} depicts the correlation between $u_{\alpha, i}(T)$ and the relative number of interest-aligned videos $c_{\alpha, i}(T)$.}  

    \label{fig:exploration}
\end{figure}

Specifically, we examined the number of unique hashtags $H_u(t)$ normalized by the total number of hashtags $H_T(t)$ encountered during the experiments, defined as $u(t) = H_u(t)/H_T(t)$.  
While the accumulation of unique hashtags over time can be expected to follow patterns characteristic of real-world discovery processes, such as Heaps' or Zipf's law \cite{iacopini_interacting_2020,heaps1978information,zipf2016human}, the total number of encountered hashtags will grow linearly at a rate proportional to the mean number of hashtags per video.  
The interaction of these two growth dynamics results in the type of power-law decay observed in $u(t)$ {for each experimental realization}, as shown in Fig.~\ref{fig:exploration}A.  
In this figure, colors represent experimental conditions, thin lines correspond to individual experimental realizations, and thick lines indicate the average across realizations within each condition.  

Consistent with the assumption of topic-specific content dynamics, we find that the degree of exploration depends on the bots' interests: bots interested in \textsc{Food} explore a greater number of unique hashtags than those focused on \textsc{Gaming}.  
The exploration dynamics of double-interest bots (\textsc{Gaming+Food}) generally fall between those of the two single-interest bots.  
This contrast in content exploration is further highlighted in Fig.~\ref{fig:exploration}B, which depicts the final ratios of unique to total hashtag counts, $u(t=T)$, across experimental conditions.

Finally, how does content exploration relate to the extent to which the feed algorithm aligns with the bots' interests?  
To address this question, we examine the correlation between the normalized cumulative count of interest-aligned content, $c(T)=C(T)/T$, where $T$ is the total number of videos the bot has encountered, and the normalized ratio of unique hashtags, $u(T)$.  
As shown in Fig.~\ref{fig:exploration}C, these two quantities exhibit strong negative correlations across all three conditions.  
This correlation is strongest for bots interested in \textsc{Gaming} ($r=-0.92$) and weakest for those interested in \textsc{Gaming+Food} ($r=-0.71$), with \textsc{Food} bots falling in between ($r=-0.8$).  

\section{Discussion}  
In the digital age, investigating how users' interests influence the amplification of content on online platforms is crucial to understand the dynamics of today's information ecosystem and its influence on human culture \cite{brinkmann_machine_2023}. {Addressing this issue on a platform like TikTok requires a user-centric methodology, which we implemented through a sock-puppet audit. Here, bot behavior is designed to simulate human behavior online, leveraging a large language model to evaluate and respond to content in real-time.}
 
{Our findings reveal that TikTok's algorithmic feed strongly amplifies content aligned with user interests.}
Both single- and dual-interest bots experienced significant increases in the proportion of interest-matching content over time, with amplification factors varying across experimental conditions. 
{Typically}, the onset of content amplification occurs rapidly, often within the first 200 videos, with bots interested in \textsc{Gaming} content experiencing the fastest onset and \textsc{Food} bots the slowest. 

{After this initial phase, the recommendation dynamics exhibit distinct patterns that we uncovered and formalized using Markov models.}
First, simple Markov models reveal consistently elevated transition probabilities toward interest-aligned content across all conditions.  %Second, hidden Markov models uncover significant differences in the complexity of the recommendation dynamics across interests, i.e. the number of hidden states and the transitions between them {across bots' interests}.
Second, hidden Markov models reveal significant differences in the complexity of recommendation dynamics across bot interests, specifically in the number of hidden states and the transitions between them. 
We also find that interest-aligned content is, on average, less popular (lower like counts) and longer in duration than non-interest content, suggesting a potential trade-off between personalization and mainstream appeal. Furthermore, content exploration, as measured by the number of unique hashtags encountered by the bots, exhibits a strong negative correlation with content amplification: bots receiving more interest-aligned content experience lower diversity in hashtag exposure.

Our findings build on existing research on algorithmic feeds and their role in content amplification.  
{Unlike previous studies using sock-puppet audits \cite{boeker_empirical_2022,tjaden_automated_2024}, we found that TikTok’s feed algorithm not only strongly amplifies content aligned with a user's interests but does so rapidly, typically within approximately 200 videos—equivalent to roughly 1,5 hours of browsing.}
This rapid and pronounced content amplification raises concerns about algorithmic feedback loops, where initial user signals become increasingly magnified over time, potentially limiting exposure to diverse perspectives \cite{piao_humanai_2023,cinelli_echo_2021}.
Notably, the varying amplification strengths across content types suggest domain-specific differences in TikTok's feed algorithm.

Prior work on YouTube \cite{haroon_auditing_2023} and Twitter \cite{huszar_algorithmic_2022} has shown that certain topics (e.g., political content) are reinforced more strongly than others (e.g., general entertainment). Similarly, our results suggest that TikTok's algorithm introduces topic-specific biases, with \textsc{Gaming} content undergoing greater amplification than \textsc{Food} content.  
A key question arising from our results is why algorithms like TikTok’s "For You" feed exhibits such differential amplification across user interests. Several factors could contribute to this phenomenon.  
First, engagement-driven reinforcement may play a role, where higher user engagement (e.g., watch time, likes, shares) for certain content categories prompts the algorithm to prioritize these topics more aggressively.  
{Second, differences in content supply dynamics may influence amplification.  
The prevalence of certain content types on the platform can shape the extent to which they are reinforced. Less frequently appearing content is less likely to be encountered by users, thereby reducing its potential for amplification.  
}
{Finally, algorithmic learning rates may vary across content types, possibly due to variations in content classification accuracy or ranking mechanisms. \textsc{Gaming} content, for instance, may be more narrowly defined and easier to classify, which could contribute to its stronger amplification. This raises broader concerns about the amplification of other niche content, including potentially harmful material such as radical political videos, which could facilitate the formation of echo chambers and reinforce ideological polarization online  
\cite{baumann_modeling_2020,baumann_emergence_2021,cinelli_echo_2021}.  
}

Despite the strong amplification of interest-aligned content, exploration remains substantial. 
The stationary content distributions inferred from the Markov models indicate that even in the strongest cases of amplification, roughly one-third of the content remains unaligned with a user's interests.  
This suggests that TikTok's feed algorithm indeed balances personalization and content diversity, prioritizing content alignment while preventing complete convergence.  
Additionally, the similar stationary probabilities observed between single- and dual-interest bots indicate that the algorithm does not necessarily amplify content more aggressively for users with fewer interests. Importantly, we find strong negative correlations between the amount of interest-aligned content a bot engages with during an experiment and the extent of its exploration within the hashtag space. These correlations suggest a potential downside of algorithmic amplification: as user interests become increasingly reinforced, exposure to new content on the platform may be restricted, thereby limiting content diversity on the level of individuals \cite{baumann_optimal_2024}.

Several limitations of this study should be acknowledged.  
While bots provide controlled insights into recommendation dynamics, they are nowhere near to perfectly replicate real user behavior, particularly in terms of {more heterogeneous} interactions with the platform. 
Each time a bot encountered interest-aligned content, it consistently liked and rewatched the video while also following its creator, limiting the variability of engagement patterns observed in human users.  
Additionally, our bots did not analyze video content directly but instead inferred the topic of the video based on textual metadata, i.e., the video descriptions and hashtags. 
This reliance on textual cues introduces uncertainties in accurately identifying interest-aligned content.  
{Another challenge in quantifying content amplification in our experiments is the absence of a "neutral baseline," i.e., bots \textit{without} predefined interests and the content they would encounter over time. Due to TikTok's bot detection system, we were unable to collect such data, as interest-neutral bots were removed from the platform during the experiments.}
Most importantly, our study focused on only two content domains (\textsc{Gaming} and \textsc{Food}), which may not generalize to other topics. 
Future research should extend this approach to politically sensitive or controversial content to examine whether similar amplification patterns emerge, {and how those pattern differ across the globe, as we have focused on bots residing in the US.}
Finally, while our experiments covers an extended period of roughly one week per bot, even longer-term studies could offer deeper insights into whether amplification actually plateaus continues to escalate or oscillates over time.  

This study provides a detailed empirical analysis of TikTok’s recommendation dynamics, highlighting the rapid onset and sustained amplification of interest-aligned content. While strong content reinforcement occurs, TikTok's "For You" feed maintains a level of exploration that prevents complete collapse of content. The differential amplification across content types, the trade-off between personalization and content popularity, and the distinct phases of recommendation dynamics revealed by Hidden Markov Models provide important insights into the mechanisms shaping algorithmic curation and contribute to ongoing discussions about the role of recommender systems in shaping digital information environments.

\section{Methods}  

\subsection{Sock-Puppet Audit Design}  
To investigate TikTok’s recommendation algorithm, we conducted a sock-puppet audit, deploying automated bots to interact with the platform's content feed under controlled conditions. Given TikTok’s advanced bot detection mechanisms, we utilized Selenium as the primary tool for automation, supplemented with \textit{Selenium Stealth} to minimize detectability and ensure that the bots' actions closely mimicked human-like behavior.

\subsection{Experimental Setup}  
All experiments were conducted within American time zones (9 AM to 2 AM EST) to align with realistic user activity patterns. To ensure exposure to U.S.-based content, we employed proxies to spoof the bots' locations to American servers.  

Each bot was assigned one of three predefined interests: (i) \textsc{Gaming}, (ii) \textsc{Food}, or (iii) \textsc{Gaming+Food} (dual-interest). While initial tests explored a broader range of topics, these categories were selected due to their prevalence on TikTok, ensuring sufficient data collection for robust analysis.

{Each experimental realization (i.e., each bot) spanned multiple days of TikTok browsing, with bots watching approximately 500 videos per day. To minimize the risk of detection due to repetitive behavior, session lengths varied, and bots did not watch all 500 videos in a single sitting. Instead, they logged onto the platform for sessions of approximately 100 videos ($\pm10$).}

{After each session, bots saved cookies, logged out, and remained inactive for a randomized period of 20--30 minutes before initiating the subsequent session. Each bot was associated with a distinct \textit{Chrome profile}, allowing automatic re-authentication via stored credentials. This setup ensured seamless login, thereby circumventing TikTok's CAPTCHA verification.}

{Unless an earlier termination was required for other reasons, each bot was run for approximately one week (i.e., 7 days), resulting in over 3,000 videos watched in most cases.}

\subsection{Bot Behavior and Interaction}
{Bots were created manually, each assigned a unique name and email address. To generate names for TikTok accounts, we randomly selected from a dataset provided by the Python \texttt{name} API. 
All bots were registered with a male gender designation, as the influence of gender on content amplification was not the focus of this study.
A complete list of experimental realizations and associated statistics is provided in the SM.}

During each session, bots logged into TikTok, viewed video streams, and engaged in user-like interactions based on the encountered content. Engagement decisions were guided by a language model (\textit{GPT-3.5 Turbo}), which evaluated video descriptions and hashtags to determine their relevance to the bot’s assigned interests. To maintain consistency with an English-speaking user profile, videos in languages other than English were skipped. The specific prompts provided to GPT-3.5 Turbo for content evaluation and language identification are detailed in the SM.

If a video was deemed relevant, the bot engaged with it by performing the following three actions: (i) liking the video, {(ii) rewatching it a randomly determined number of times between 1 and 3}, and (iii) following the content creator. Conversely, for non-relevant videos, the bot minimized engagement by watching only 5–10\% of the total video duration before scrolling to the next video.  

\subsection{Data Collection and Storage}  
The study was conducted over three months [June ('2024-06-04 15:00:53')–August ('2024-08-18 18:52:55')], during which seven test cycles were completed, deploying a total of 42 bots. Each test lasted approximately 6–7 days, with each bot watching an average of 3,000 videos per test cycle.

In total, the bots processed 128,148 videos, of which 66,672 were unique.  
For analysis, we excluded videos lacking descriptions or hashtags, resulting in a final dataset of 112,610 videos, including 58,213 unique entries.  
The cleaned dataset contained a total of 817,200 hashtags, of which 63,694 were unique.  
Additionally, the dataset included 27,507 distinct content creators, with an average video duration of 79 seconds.  
On average, bots watched approximately 64\% of each video's total length.  

All collected data, including video-specific details such as duration, description, and hashtags, along with additional metadata such as the content creator's name and the number of likes, were stored in a \textit{MongoDB Atlas} database for subsequent analysis.  

\subsection{Reproducibility and Code Availability}  
To facilitate replication of our results and provide transparency regarding bot creation and deployment, we will include detailed setup instructions in the README file of the project repository. The repository will be made available at:  

\textbf{\texttt{https://gitlab.gwdg.de/arora/tiktok-audit}}  

This documentation will outline the necessary steps to create and configure the bots and execute  experiments on TikTok.

\textbf{Disclaimer:} TikTok’s platform and underlying feed algorithm are frequently updated, which may impact the functionality of our sock-puppet auditing framework. As the code is not actively maintained, it may become outdated at the time of reproduction, requiring modifications to ensure compatibility with the platform's latest version.

\subsection*{Acknowledgements}
We thank Ángel Cuevas Rumín for his insightful comments on the implementation of the sock puppet audit.

\bibliography{scibib}
\bibliographystyle{Science}

% Now include supplementary material
\clearpage
\appendix
\input{SM.tex}  % This merges SM.tex content here

% Following is a new environment, {scilastnote}, that's defined in the
% preamble and that allows authors to add a reference at the end of the
% list that's not signaled in the text; such references are used in
% *Science* for acknowledgments of funding, help, etc.

% For your review copy (i.e., the file you initially send in for
% evaluation), you can use the {figure} environment and the
% \includegraphics command to stream your figures into the text, placing
% all figures at the end.  For the final, revised manuscript for
% acceptance and production, however, PostScript or other graphics
% should not be streamed into your compliled file.  Instead, set
% captions as simple paragraphs (with a \noindent tag), setting them
% off from the rest of the text with a \clearpage as shown  below, and
% submit figures as separate files according to the Art Department's
% instructions.

\end{document}

%% file: SM.tex
\section*{Supplemental Material of\\ \textit{Dynamics of Algorithmic Content Amplification on TikTok}}
\subsection*{Markov modeling}
\subsubsection*{Simple Markov models}
We implemented the Markov models using Python with NumPy. Transition matrices were computed by iterating through observed state transitions in the time series data, normalizing row-wise to represent probabilities. The stationary distributions were derived from the eigenvectors of the transition matrices associated with eigenvalue 1. Complete code implementation and parameters are available in the project repository published alongside the manuscript.

\subsubsection*{Hidden Markov models}
We implemented Hidden Markov Models (HMMs) to analyze recommendation dynamics beyond observable state transitions. The models were trained using the \textit{hmmlearn} library in Python. Observations were binary signals derived from bot interactions with TikTok content. Model selection employed Bayesian Information Criterion (BIC) scores, evaluating models with varying numbers of hidden states and multiple random initializations to identify the best fit. Detailed procedures and complete implementation details are available in the associated project repository.

\subsection*{Data Description}
This section provides a detailed overview of the data collected during the sock-puppet audit. The tables \ref{tab:videos_stats_gaming}-\ref{tab:videos_stats_gaming_food} summarize key statistics for each bot, including the total number of videos watched, the proportion of interest-aligned content encountered, and other relevant metrics. These datasets form the basis for our analysis of TikTok’s recommendation dynamics, allowing us to systematically compare content amplification patterns across different experimental conditions.

%tab:videos_stats_gaming
\begin{table}[]
\caption{\textsc{Gaming} condition.}
\label{tab:videos_stats_gaming}
\resizebox{\columnwidth}{!}{%
\begin{tabular}{|l|l|l|l|l|l|l|l|l|l|l}
\cline{1-10}
\begin{tabular}[c]{@{}l@{}}run\\ id\end{tabular} & \begin{tabular}[c]{@{}l@{}}number \\ of videos\end{tabular} & \begin{tabular}[c]{@{}l@{}}number (percent) \\ of GAMING videos\end{tabular} & \begin{tabular}[c]{@{}l@{}}number (percent) \\ of FOOD videos\end{tabular} & \begin{tabular}[c]{@{}l@{}}number of \\ hashtags\end{tabular} & \begin{tabular}[c]{@{}l@{}}number \\ of unique \\ hashtags\end{tabular} & start date & end date   & \begin{tabular}[c]{@{}l@{}}number \\ of batches\end{tabular} & \begin{tabular}[c]{@{}l@{}}average number \\ of videos \\ per batch\end{tabular} &  \\ \cline{1-10}
1                                                & 1835                                                        & 973 (53.0\%)                                                                 & 50 (2.7\%)                                                                 & 14634                                                         & 4388                                                                    & 2024-06-04 & 2024-06-09 & 28                                                           & 65.5                                                                             &  \\ \cline{1-10}
2                                                & 2468                                                        & 1362 (55.2\%)                                                                & 88 (3.6\%)                                                                 & 19875                                                         & 5518                                                                    & 2024-06-04 & 2024-06-09 & 30                                                           & 82.3                                                                             &  \\ \cline{1-10}
3                                                & 2813                                                        & 1547 (55.0\%)                                                                & 71 (2.5\%)                                                                 & 21453                                                         & 5533                                                                    & 2024-06-12 & 2024-06-20 & 36                                                           & 78.1                                                                             &  \\ \cline{1-10}
4                                                & 1676                                                        & 1106 (66.0\%)                                                                & 44 (2.6\%)                                                                 & 14498                                                         & 4046                                                                    & 2024-06-12 & 2024-06-20 & 31                                                           & 54.1                                                                             &  \\ \cline{1-10}
5                                                & 3031                                                        & 1965 (64.8\%)                                                                & 71 (2.3\%)                                                                 & 24748                                                         & 6826                                                                    & 2024-06-28 & 2024-07-05 & 35                                                           & 86.6                                                                             &  \\ \cline{1-10}
6                                                & 2953                                                        & 1880 (63.7\%)                                                                & 69 (2.3\%)                                                                 & 22376                                                         & 6435                                                                    & 2024-06-28 & 2024-07-05 & 35                                                           & 84.4                                                                             &  \\ \cline{1-10}
7                                                & 3190                                                        & 2433 (76.3\%)                                                                & 53 (1.7\%)                                                                 & 23985                                                         & 6647                                                                    & 2024-07-07 & 2024-07-14 & 38                                                           & 83.9                                                                             &  \\ \cline{1-10}
8                                                & 3245                                                        & 2970 (91.5\%)                                                                & 23 (0.7\%)                                                                 & 21178                                                         & 3296                                                                    & 2024-07-07 & 2024-07-15 & 38                                                           & 85.4                                                                             &  \\ \cline{1-10}
9                                                & 534                                                         & 453 (84.8\%)                                                                 & 3 (0.6\%)                                                                  & 4867                                                          & 1062                                                                    & 2024-07-16 & 2024-07-18 & 7                                                            & 76.3                                                                             &  \\ \cline{1-10}
10                                               & 2284                                                        & 2100 (91.9\%)                                                                & 15 (0.7\%)                                                                 & 15978                                                         & 2387                                                                    & 2024-07-16 & 2024-07-23 & 30                                                           & 76.1                                                                             &  \\ \cline{1-10}
11                                               & 3423                                                        & 3110 (90.9\%)                                                                & 34 (1.0\%)                                                                 & 21714                                                         & 3846                                                                    & 2024-07-26 & 2024-08-05 & 40                                                           & 85.6                                                                             &  \\ \cline{1-10}
12                                               & 3585                                                        & 3199 (89.2\%)                                                                & 25 (0.7\%)                                                                 & 25480                                                         & 4246                                                                    & 2024-07-26 & 2024-08-05 & 40                                                           & 89.6                                                                             &  \\ \cline{1-10}
13                                               & 2384                                                        & 2034 (85.3\%)                                                                & 24 (1.0\%)                                                                 & 18226                                                         & 3830                                                                    & 2024-08-09 & 2024-08-17 & 31                                                           & 76.9                                                                             &  \\ \cline{1-10}
14                                               & 2720                                                        & 2117 (77.8\%)                                                                & 54 (2.0\%)                                                                 & 21879                                                         & 5157                                                                    & 2024-08-09 & 2024-08-18 & 34                                                           & 80.0                                                                             &  \\ \cline{1-10}
\end{tabular}%
}
\end{table}

%tab:videos_stats_food
\begin{table}[]
\caption{\textsc{Food} condition.}
\label{tab:videos_stats_food}
\resizebox{\columnwidth}{!}{%
\begin{tabular}{|l|l|l|l|l|l|l|l|l|l|l}
\cline{1-10}
\begin{tabular}[c]{@{}l@{}}run\\ id\end{tabular} & \begin{tabular}[c]{@{}l@{}}number \\ of videos\end{tabular} & \begin{tabular}[c]{@{}l@{}}number (percent) \\ of GAMING videos\end{tabular} & \begin{tabular}[c]{@{}l@{}}number (percent) \\ of FOOD videos\end{tabular} & \begin{tabular}[c]{@{}l@{}}number of \\ hashtags\end{tabular} & \begin{tabular}[c]{@{}l@{}}number \\ of unique \\ hashtags\end{tabular} & start date & end date   & \begin{tabular}[c]{@{}l@{}}number \\ of batches\end{tabular} & \begin{tabular}[c]{@{}l@{}}average number \\ of videos \\ per batch\end{tabular} &  \\ \cline{1-10}
1                                                & 2163                                                        & 221 (10.2\%)                                                                 & 789 (36.5\%)                                                               & 14859                                                         & 5647                                                                    & 2024-06-04 & 2024-06-09 & 29                                                           & 74.6                                                                             &  \\ \cline{1-10}
2                                                & 2220                                                        & 184 (8.3\%)                                                                  & 842 (37.9\%)                                                               & 14790                                                         & 5691                                                                    & 2024-06-04 & 2024-06-09 & 30                                                           & 74.0                                                                             &  \\ \cline{1-10}
3                                                & 2866                                                        & 114 (4.0\%)                                                                  & 1442 (50.3\%)                                                              & 19229                                                         & 7244                                                                    & 2024-06-12 & 2024-06-20 & 40                                                           & 71.7                                                                             &  \\ \cline{1-10}
4                                                & 2106                                                        & 18 (0.9\%)                                                                   & 1573 (74.7\%)                                                              & 14631                                                         & 3160                                                                    & 2024-06-12 & 2024-06-20 & 26                                                           & 81.0                                                                             &  \\ \cline{1-10}
5                                                & 2793                                                        & 88 (3.2\%)                                                                   & 1826 (65.4\%)                                                              & 23203                                                         & 7183                                                                    & 2024-06-28 & 2024-07-06 & 35                                                           & 79.8                                                                             &  \\ \cline{1-10}
6                                                & 2712                                                        & 76 (2.8\%)                                                                   & 1318 (48.6\%)                                                              & 18372                                                         & 6800                                                                    & 2024-06-28 & 2024-07-05 & 35                                                           & 77.5                                                                             &  \\ \cline{1-10}
7                                                & 2620                                                        & 19 (0.7\%)                                                                   & 2240 (85.5\%)                                                              & 20744                                                         & 5669                                                                    & 2024-07-07 & 2024-07-15 & 32                                                           & 81.9                                                                             &  \\ \cline{1-10}
8                                                & 3265                                                        & 129 (4.0\%)                                                                  & 2039 (62.5\%)                                                              & 27018                                                         & 7718                                                                    & 2024-07-07 & 2024-07-15 & 44                                                           & 74.2                                                                             &  \\ \cline{1-10}
9                                                & 2584                                                        & 81 (3.1\%)                                                                   & 1536 (59.4\%)                                                              & 19167                                                         & 6606                                                                    & 2024-07-16 & 2024-07-24 & 38                                                           & 68.0                                                                             &  \\ \cline{1-10}
10                                               & 2451                                                        & 52 (2.1\%)                                                                   & 1742 (71.1\%)                                                              & 16105                                                         & 6022                                                                    & 2024-07-16 & 2024-07-24 & 35                                                           & 70.0                                                                             &  \\ \cline{1-10}
11                                               & 2862                                                        & 17 (0.6\%)                                                                   & 2423 (84.7\%)                                                              & 20519                                                         & 6032                                                                    & 2024-07-26 & 2024-08-06 & 35                                                           & 81.8                                                                             &  \\ \cline{1-10}
12                                               & 3144                                                        & 19 (0.6\%)                                                                   & 2588 (82.3\%)                                                              & 23334                                                         & 7078                                                                    & 2024-07-26 & 2024-08-05 & 40                                                           & 78.6                                                                             &  \\ \cline{1-10}
13                                               & 2530                                                        & 42 (1.7\%)                                                                   & 1677 (66.3\%)                                                              & 18700                                                         & 6196                                                                    & 2024-08-09 & 2024-08-18 & 40                                                           & 63.2                                                                             &  \\ \cline{1-10}
14                                               & 1692                                                        & 36 (2.1\%)                                                                   & 1078 (63.7\%)                                                              & 12897                                                         & 4981                                                                    & 2024-08-09 & 2024-08-18 & 24                                                           & 70.5                                                                             &  \\ \cline{1-10}
\end{tabular}%
}
\end{table}

%tab:videos_stats_gaming_food
\begin{table}[]
\caption{\textsc{Gaming+Food} condition.}
\label{tab:videos_stats_gaming_food}
\resizebox{\columnwidth}{!}{%
\begin{tabular}{|l|l|l|l|l|l|l|l|l|l|l}
\cline{1-10}
\begin{tabular}[c]{@{}l@{}}run\\ id\end{tabular} & \begin{tabular}[c]{@{}l@{}}number \\ of videos\end{tabular} & \begin{tabular}[c]{@{}l@{}}number (percent) \\ of GAMING videos\end{tabular} & \begin{tabular}[c]{@{}l@{}}number (percent) \\ of FOOD videos\end{tabular} & \begin{tabular}[c]{@{}l@{}}number of \\ hashtags\end{tabular} & \begin{tabular}[c]{@{}l@{}}number \\ of unique \\ hashtags\end{tabular} & start date & end date   & \begin{tabular}[c]{@{}l@{}}number \\ of batches\end{tabular} & \begin{tabular}[c]{@{}l@{}}average number \\ of videos \\ per batch\end{tabular} &  \\ \cline{1-10}
1                                                & 1518                                                        & 676 (44.5\%)                                                                 & 164 (10.8\%)                                                               & 13068                                                         & 4162                                                                    & 2024-06-04 & 2024-06-10 & 19                                                           & 79.9                                                                             &  \\ \cline{1-10}
2                                                & 1932                                                        & 782 (40.5\%)                                                                 & 167 (8.6\%)                                                                & 15658                                                         & 5239                                                                    & 2024-06-04 & 2024-06-09 & 25                                                           & 77.3                                                                             &  \\ \cline{1-10}
3                                                & 2294                                                        & 942 (41.1\%)                                                                 & 571 (24.9\%)                                                               & 17585                                                         & 5335                                                                    & 2024-06-12 & 2024-06-20 & 34                                                           & 67.5                                                                             &  \\ \cline{1-10}
4                                                & 2573                                                        & 1226 (47.6\%)                                                                & 346 (13.4\%)                                                               & 18729                                                         & 5659                                                                    & 2024-06-12 & 2024-06-20 & 36                                                           & 71.5                                                                             &  \\ \cline{1-10}
5                                                & 3117                                                        & 2036 (65.3\%)                                                                & 353 (11.3\%)                                                               & 25807                                                         & 6187                                                                    & 2024-06-28 & 2024-07-06 & 35                                                           & 89.1                                                                             &  \\ \cline{1-10}
6                                                & 3108                                                        & 2072 (66.7\%)                                                                & 372 (12.0\%)                                                               & 22300                                                         & 5523                                                                    & 2024-06-28 & 2024-07-06 & 35                                                           & 88.8                                                                             &  \\ \cline{1-10}
7                                                & 3085                                                        & 801 (26.0\%)                                                                 & 1863 (60.4\%)                                                              & 22942                                                         & 5997                                                                    & 2024-07-07 & 2024-07-15 & 39                                                           & 79.1                                                                             &  \\ \cline{1-10}
8                                                & 3143                                                        & 2862 (91.1\%)                                                                & 68 (2.2\%)                                                                 & 20609                                                         & 3422                                                                    & 2024-07-07 & 2024-07-15 & 38                                                           & 82.7                                                                             &  \\ \cline{1-10}
9                                                & 1774                                                        & 10 (0.6\%)                                                                   & 1438 (81.1\%)                                                              & 12822                                                         & 3996                                                                    & 2024-07-16 & 2024-07-23 & 25                                                           & 71.0                                                                             &  \\ \cline{1-10}
10                                               & 2619                                                        & 2074 (79.2\%)                                                                & 28 (1.1\%)                                                                 & 18908                                                         & 3553                                                                    & 2024-07-16 & 2024-07-24 & 34                                                           & 77.0                                                                             &  \\ \cline{1-10}
11                                               & 3560                                                        & 2974 (83.5\%)                                                                & 295 (8.3\%)                                                                & 24979                                                         & 4280                                                                    & 2024-07-26 & 2024-08-05 & 40                                                           & 89.0                                                                             &  \\ \cline{1-10}
12                                               & 3159                                                        & 2243 (71.0\%)                                                                & 518 (16.4\%)                                                               & 22237                                                         & 4684                                                                    & 2024-07-26 & 2024-08-05 & 40                                                           & 79.0                                                                             &  \\ \cline{1-10}
13                                               & 2472                                                        & 1778 (71.9\%)                                                                & 180 (7.3\%)                                                                & 21966                                                         & 4536                                                                    & 2024-08-09 & 2024-08-18 & 34                                                           & 72.7                                                                             &  \\ \cline{1-10}
14                                               & 2488                                                        & 214 (8.6\%)                                                                  & 1555 (62.5\%)                                                              & 18337                                                         & 5976                                                                    & 2024-08-09 & 2024-08-18 & 34                                                           & 73.2                                                                             &  \\ \cline{1-10}
\end{tabular}%
}
\end{table}

\subsection*{Change point detection and curve fitting}

This section provides additional details on the methods used to identify key transitions in TikTok’s recommendation dynamics. To quantify the onset of content amplification, we apply change point detection techniques to the time series of interest-aligned content rates. This allows us to determine when TikTok’s feed algorithm begins significantly reinforcing user interests. 

Following this, we employ curve fitting to characterize the amplification trends, estimating the rate at which interest-aligned content increases over time. These analyses complement the results presented in the main text by offering a more precise quantification of the amplification dynamics and its variation across experimental conditions.

\begin{figure}[!h]
    \centering
    \includegraphics[width=\linewidth]{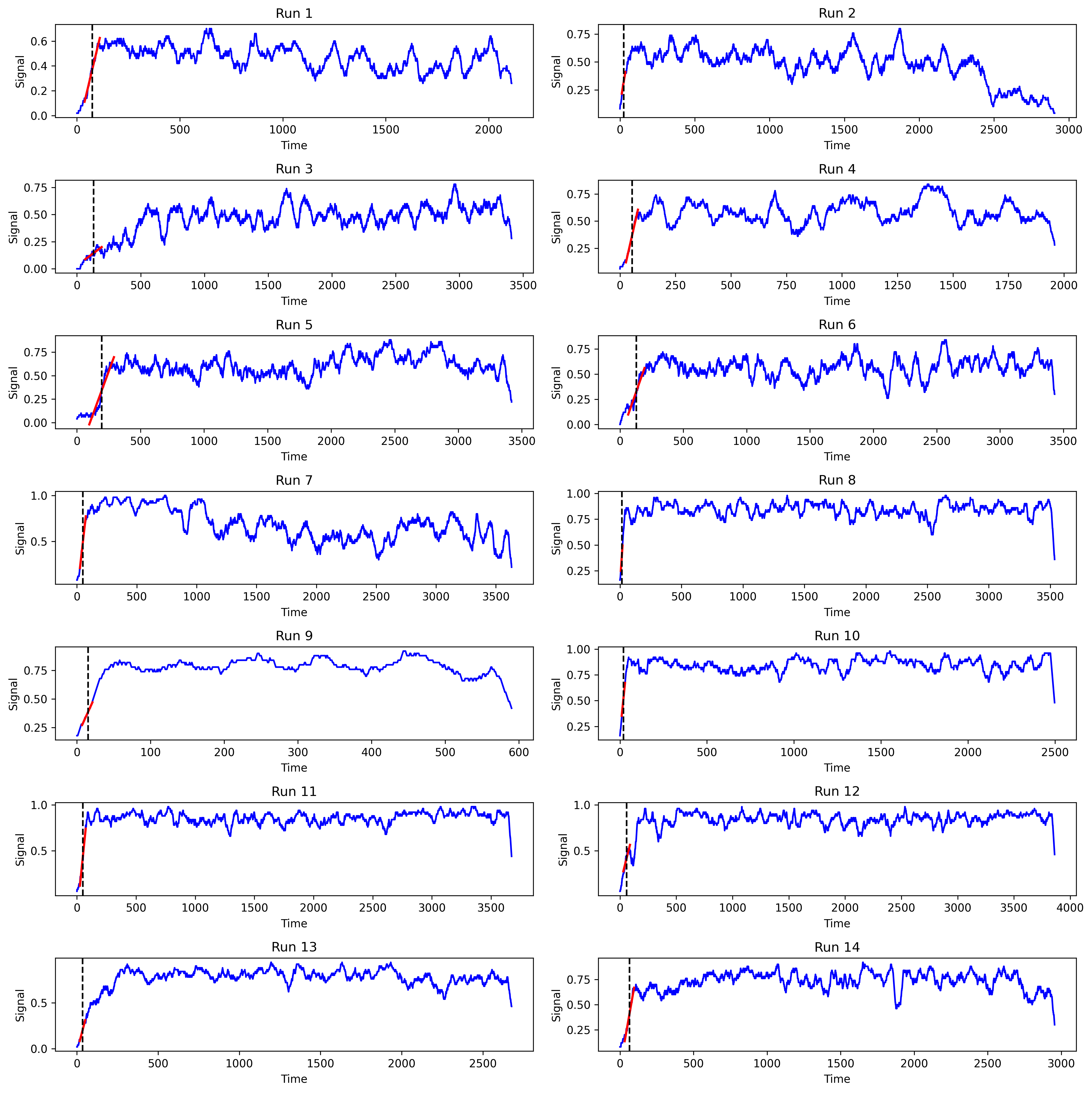}
    \caption{\textsc{Gaming} condition.}
\label{fig:change_points_gaming}
\end{figure}

\begin{figure}[!h]
    \centering
    \includegraphics[width=\linewidth]{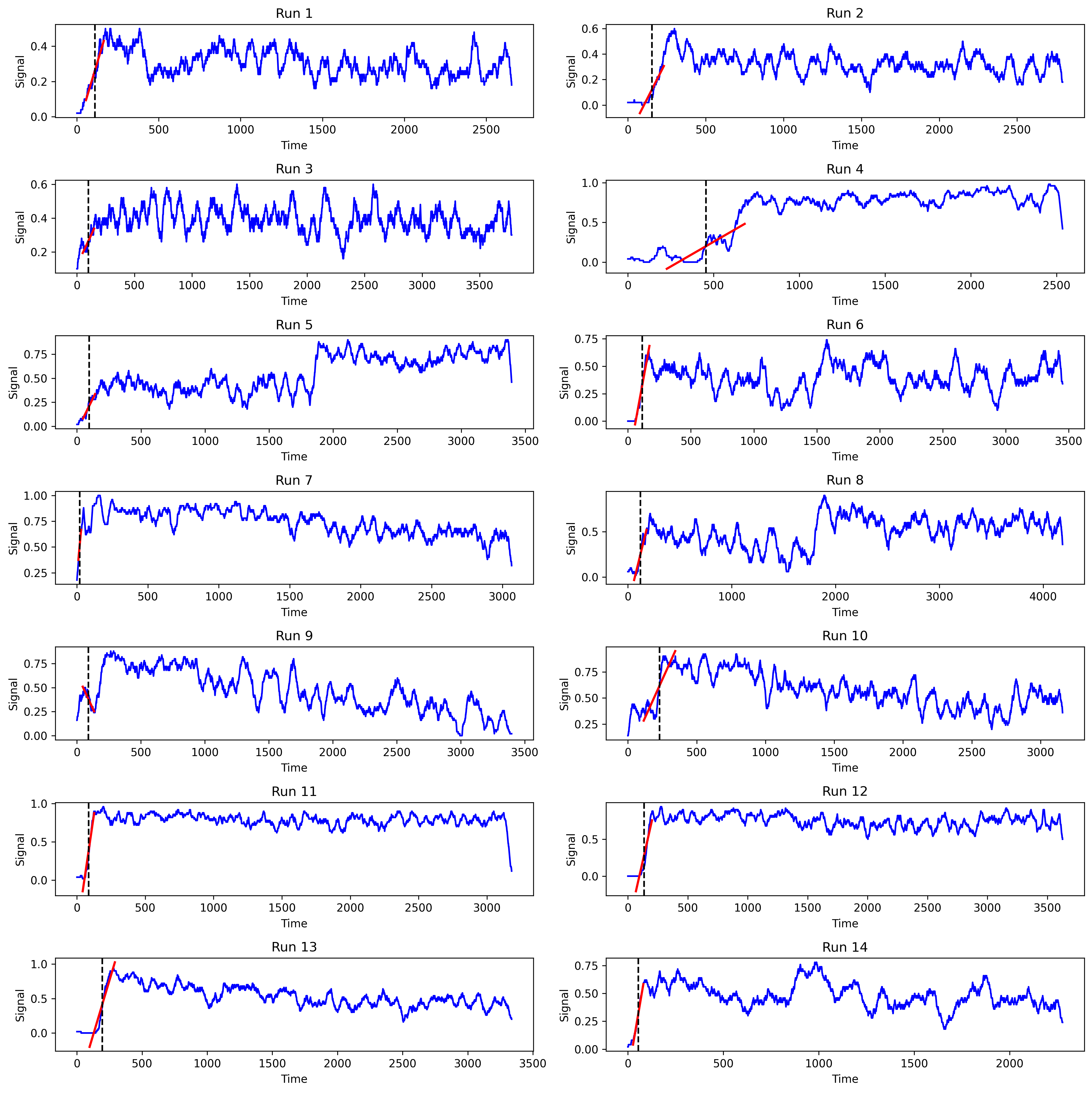}
    \caption{\textsc{Food} condition.}
\label{fig:change_points_food}
\end{figure}

\begin{figure}[!h]
    \centering
    \includegraphics[width=\linewidth]{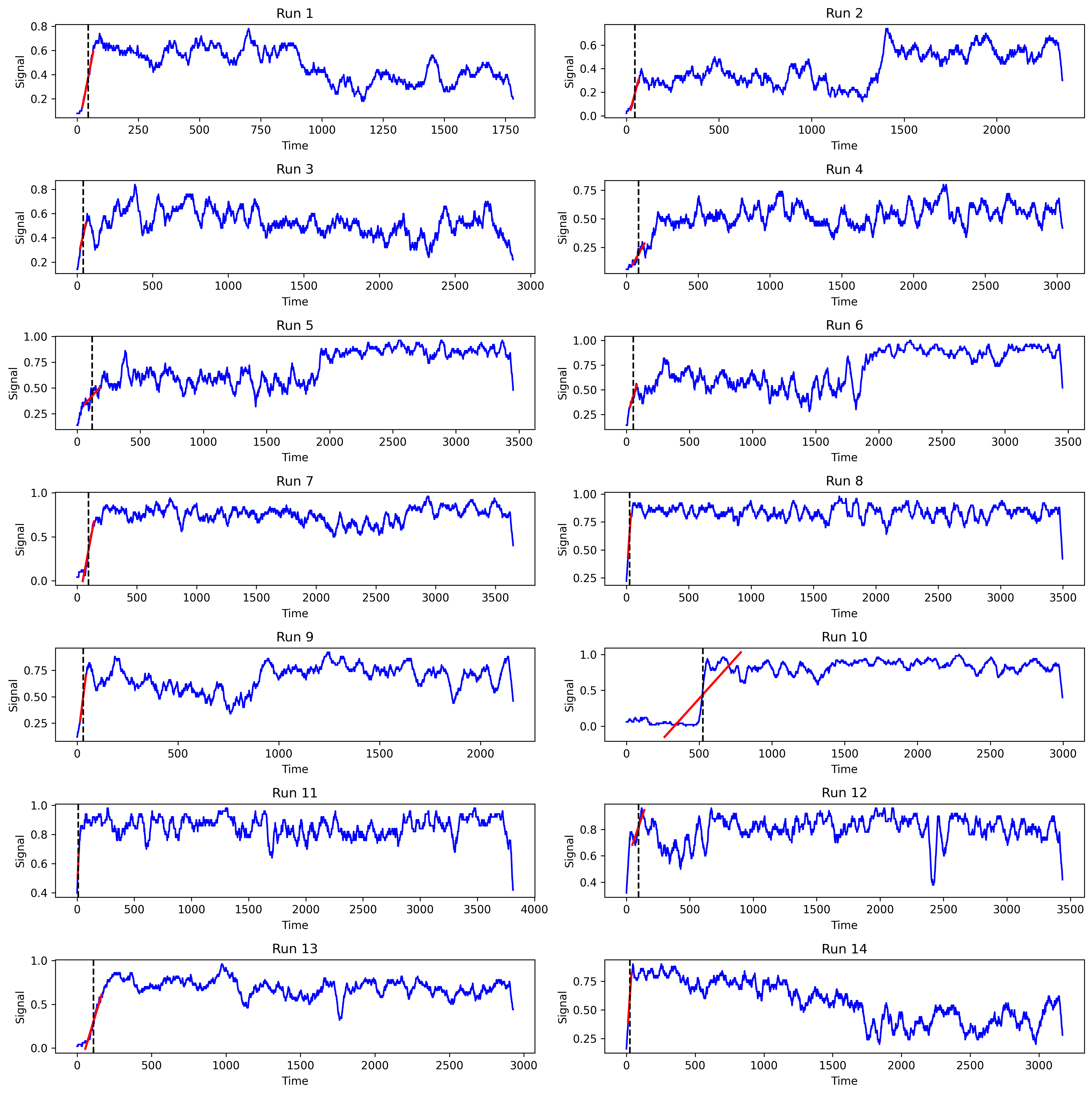}
    \caption{\textsc{Gaming}+\textsc{Food} condition.}
\label{fig:change_points_gaming_food}
\end{figure}

\subsection*{Details on GPT-3.5 Turbo prompts}

Engagement decisions in our sock-puppet audit were determined using the \textit{GPT-3.5 Turbo} model (exact model: \texttt{gpt-35-turbo}, API version: \texttt{2024-02-15-preview}) provided by OpenAI. Specifically, we employed two distinct prompts for each video encountered by the bots.

First, to verify that the content was in English, we used the following prompt:
\begin{quote}
\texttt{I have a video description: "\{description\}" and hashtags: "\{hashtags\}" extracted from a video. Classify whether the information is in English or not. Give a True if the information holds a semantic meaning in English else give False.}
\end{quote}

Second, to determine if the content matched the bot's assigned topic, we provided the following prompt:
\begin{quote}
\texttt{I have a video description: "\{description\}" and hashtags: "\{hashtags\}" extracted from a video. Classify whether the information is related to topic "\{topic\}". Give a True if the information is related to \{topic\} else give False.}
\end{quote}

Based on the model’s responses, bots either engaged with or skipped the video content.

%\bibliography{scibib}
%\bibliographystyle{Science}

% \usepackage{graphicx}

% Following is a new environment, {scilastnote}, that's defined in the
% preamble and that allows authors to add a reference at the end of the
% list that's not signaled in the text; such references are used in
% *Science* for acknowledgments of funding, help, etc.

% For your review copy (i.e., the file you initially send in for
% evaluation), you can use the {figure} environment and the
% \includegraphics command to stream your figures into the text, placing
% all figures at the end.  For the final, revised manuscript for
% acceptance and production, however, PostScript or other graphics
% should not be streamed into your compliled file.  Instead, set
% captions as simple paragraphs (with a \noindent tag), setting them
% off from the rest of the text with a \clearpage as shown  below, and
% submit figures as separate files according to the Art Department's
% instructions.